\documentclass[setspace,amsmath,amssymb,fleqn]{article}
\usepackage{setspace}
\usepackage{graphicx}
\usepackage{amsmath}
\usepackage{amssymb}
\usepackage{cite}
 
\newcommand{\bx}{{\boldsymbol{x}}}

\begin{document}
\begin{titlepage}

\begin{flushright}
\today
\end{flushright}

\vspace{1in}

\begin{center}

{\bf Information-acquiring von Neumann architecture of a computer: A theoretical design}

\vspace{1in}

\normalsize

\renewcommand\thefootnote{\fnsymbol{footnote}}

{Eiji Konishi\footnote[0]{E-mail address: konishi.eiji.27c@kyoto-u.jp}
}

\normalsize
\vspace{.5in}

{\it Graduate School of Human and Environmental Studies,\\
 Kyoto University, Kyoto 606-8501, Japan}
\end{center}

\vspace{1in}

\baselineskip=24pt
\begin{abstract}
We design the information-acquiring von Neumann architecture of a computer in a fine-grained or coarse-grained model of the registers (quickly accessible memories) in the central processing unit, where information is carried by classical bits.
This architecture enables both a Hamiltonian evolution process converting a given input pure state to another output pure state of the system under consideration (functionality) and a physical process to acquire information.
The latter process is identified with the projection hypothesis (state reduction) in projective quantum measurement in the ensemble interpretation of quantum mechanics.
As a novelty of this work, we treat projective quantum measurement as a classical measurement (reduction of ignorance) in the coarse-grained model.
The main objective is to examine the present author's previously proposed state-reduction mechanism in the architecture within quantum electrodynamics in the presence of the orbital superselection rule.
As a result, the electric potential incorporated into the architecture serves as a binary switch for the state reduction.
As a consequence of this architecture, information-acquiring artificial intelligence (physically grounded induction machines or agents) can be established.

\end{abstract}

\vspace{.1in}


\vspace{.6in}

\end{titlepage}

\section{Introduction}

In this day and age, the fundamental distinction between the observer in quantum mechanics and a computer is that the former has the ability to acquire information by a physical process, whereas the latter does not.
In the ensemble interpretation of quantum mechanics, which is a framework for modern quantum measurement theory, we consider an ensemble of copies of a quantum system in {\it pure} states, which are given by state vectors, and the probabilities given by the Born rule in projective quantum measurement are reinterpreted as the statistical weights on a diagonal mixed state of state vectors in the eigenbasis of a discrete measured observable \cite{dEspagnat}.
In this interpretation of quantum mechanics, the ability to acquire information by a physical process is identified with the projection hypothesis (event reading).
The {\it projection hypothesis} selects a pure state (an eigenstate) from the diagonal mixed state (a statistical mixture of eigenstates) as obtained from another pure or mixed state by the quantum decoherence process \cite{Neumann,Luders,EPL3,IJQI}.

On the other hand, the computer has functionality.
Here, the term {\it functionality} refers to a Hamiltonian evolution process converting a given input pure state to another output pure state of the system under consideration.
In a computer with the von Neumann architecture, a program and its data are both stored in the main memory, and the physical substrate of the program run on this computer is reduced to the arithmetic logic unit (ALU) in the central processing unit (CPU) \cite{NS}.
Informatically, we identify the two 1-bit states in the ALU with the two 1-bit states of a pin on the CPU, respectively.
Their binary states are entangled (have statistical correlations) in a statistical mixture via their classical mechanical interactions.

A computer with the von Neumann architecture consists of a CPU and main memory, which interact with each other.\footnote{This paragraph is based on standard textbooks on computer architecture (e.g., Refs. \cite{NS,PH}).}
The CPU consists of three parts: an ALU, registers (quickly accessible memories) with addresses in a file, and a control unit.
Here, the control unit fetches an instruction memory, stores it in the instruction register, decodes it, and executes its operation (arithmetic logic operation, reading a binary datum, or writing a binary datum) using binary data stored in the registers.
Schematically, the operation by the control unit consists of three fundamental processes constituting a macro-cycle.
The first process is the transfer of a binary datum from the main memory to the registers (read operation).
The second process is the computation in the CPU.
This computation is an inner loop in which binary data are passed from the registers to the ALU, and back to the registers.
Here, the arithmetic logic operation is done by the ALU.
The third process is the transfer of a binary datum from the registers back to the main memory (write operation).

The purpose of this article is to incorporate a binary projection hypothesis into the CPU, in a fine-grained (mechanical) or coarse-grained (stochastic) model of the registers in the CPU, as its state-reduction mechanism, and then to extend the von Neumann architecture of a computer.
This incorporation of the binary projection hypothesis enables the computer to read out the results of computations performed in the CPU.
In particular, in the fine-grained and the coarse-grained models, the computer acquires zero and positive information, respectively.

We design the pins on the CPU, which are used for the input to and the output from the CPU and located between the main memory and the registers in the CPU.
We incorporate the binary projection hypothesis into them.
In our definition, a {\it pin} on the CPU is a tubule with a {\it hydrophilic membrane} as its surface and {\it liquid water} filled inside.
The diameter of this tubule is assumed to be shorter than the resonant wavelength $\approx400$ [$\mu$m] of the radiation from a water molecule between the first excited rotational state and the ground rotational state \cite{GPV}.
Namely, the surface region of the pin (a tubule) fits inside this resonance domain.
We also assume that a non-zero electric potential (an applied voltage) sloping to zero with a finite run length can propagate down the tubule on a classical mechanical time scale.
The $1$ and $0$ states of the pin represent the on and off of this electric potential, respectively.

Here, we note three comparisons between the fine-grained and the coarse-grained models of the registers in the CPU.
First, in the fine-grained model, the classical microstate is consistently pure, but in the coarse-grained model, the classical macrostate becomes a mixed state(s) owing to the stochastic nature of the time evolution.
Second, the outcome of the fine-grained model of the CPU is a unique microstate of the pins.
On the other hand, the prediction (i.e., the reduced outcome) of the coarse-grained model of the CPU is the statistics of macrostates, in other words, the statistics of microstates of the pins, as obtained from event readings: here, each macrostate is the gained knowledge.
Third, which of these two models of the CPU is selected is attributed to the sufficiency or insufficiency of the information capacity of the registers in the CPU and thus that of the data memory.
Here, we use the term {\it data memory} to refer to the region of the addresses occupied in the main memory by the data \cite{NS}.
The data memory distinguishes the knowledge of events read by the pins, stores it as information, and feeds it back to the CPU.

In this article, we use projective quantum measurement as a classical measurement (reduction of ignorance).
Essentially, this usage exists in the literature on quantum statistical mechanics (for example, see Refs. \cite{Talkner,MT}) as projective energy measurement of a quantum canonical state (a quantum thermal equilibrium state), whose statistical weights are the Boltzmann weights.
In projective quantum measurement of a non-trivial quantum superposition (a quantum pure state), after the complete quantum decoherence (for its criterion, see Ref. \cite{JSTAT2}), there exists no unique pure state: this is the nature of quantum mechanics.
On the other hand, in classical measurement of an isolated system, there always exists a unique pure state as the mechanical state.
This is because there are no principal uncertainties in classical mechanics.
In classical measurement, the entropy of events before an event reading has the classical statistical mechanical origin.

Throughout this article, we denote operators with a hat.

The rest of this article is organized as follows.
In Sec. 2, we propose a state-reduction mechanism in a pin on the CPU based on Ref. \cite{EPL3}.
In Sec. 2.1, we detail the physical model of a pin.
In Sec. 2.2, we derive the von Neumann-type interaction, which leads to the state-reduction mechanism, from this physical model.
In Sec. 3, we conclude the article.

\section{State-Reduction Mechanism}

In this section, we propose a state-reduction mechanism in a pin on the CPU.
This mechanism is based on Ref. \cite{EPL3} and can be briefly described as follows.
For a completely decohered quantum system \cite{JSTAT2} initially in a pure or mixed state, the state reduction of its diagonal mixed state works for an event reading quantum mechanical system $\psi$ with respect to its discrete meter variable.
This is when system $\psi$ is entangled with the decohered quantum system and has the von Neumann-type interaction (see Eq. (\ref{eq:H})) \cite{dEspagnat,Neumann,IJQI} with a macroscopic and field-theoretical Bose--Einstein condensate (BEC) $A$ in the presence of the superselection rule (that is, the restriction of the complete set of the observables to an Abelian set of those that commute with the superselection rule operator) on the orbital observables of the center of mass of $A$ \cite{dEspagnat,Neumann,IJQI}.
Here, system $A$ has a non-relativistic center of mass velocity classically fluctuating in the ensemble and is initially decoupled from system $\psi$.
The field theory has spatial translational symmetry, which is broken spontaneously by the existence of a (spatiotemporally inhomogeneous) BEC \cite{EPL3}.

The core idea of this state-reduction mechanism is as follows \cite{EPL3,IJQI}.
Due to the Nambu--Goldstone theorem for spontaneously broken spatial translational symmetry, the center of mass velocity of BEC $A$ (a spatially translationally symmetric quantum mechanical system) is returned to the velocity of a c-number spatial coordinate system in the rearranged spatial coordinate system in the order parameter (i.e., the vacuum expectation value of the boson Heisenberg field) of this BEC at zero temperature \cite{Matsumoto1,Matsumoto2,Matsumoto3,Matsumoto4,Umezawa}.
Then, when system $A$ has a non-relativistic center of mass velocity classically fluctuating in the ensemble (a mixture), a constraint on the {\it state vector} (equivalent to a diagonal mixed state of the discrete meter variable with respect to the observables' statistical data \cite{JSTAT2}) of system $\psi$ arises after the von Neumann-type interaction between this system $\psi$ and system $A$, which are initially decoupled from each other.
This constraint is the physical equivalence (the Galilean relativity) between {\it state vectors} (equivalent to diagonal mixed states of the discrete meter variable with respect to the observables' statistical data \cite{JSTAT2}) with their distinct sets of a relative phase(s) of $\psi$, obtained by the partial trace of system $A$ decoupled from system $\psi$ in the mixture.
This physical equivalence is realized via the Galilean transformation of the rearranged spatial coordinate system.
This constraint is the trigger of event reading by system $\psi$.

In this mechanism, just before an event reading, the quantum state of system $\psi$ needs to be described by a {\it state vector}, which has a relative phase(s) in the superposition \cite{EPL3}.
When the projective quantum measurement is a quantum measurement of quantum coherence, the argument is straightforward \cite{EPL3}.
However, in the case of a classical measurement, the argument is novel.
When system $\psi$ is completely decohered by entangling it with the completely decohered measured quantum system \cite{JSTAT2}, system $\psi$ is in the (diagonal) mixed state.
Then, this mixed state can be purified into a {\it state vector}, without quantum coherence with respect to system $\psi$, by entangling system $\psi$ with a fictitious reference system in the form of the Schmidt decomposition \cite{NC}.

\subsection{The physical model}

In this subsection, we detail the physical model of a pin, in which the state-reduction mechanism explained above \cite{EPL3,IJQI} is to be examined.

For the state-reduction mechanism in a pin, the most important fact is that the BEC is {\it not only massive but also electrically charged} \cite{JY}.
Here, the BEC consists of evanescent (tunneling) photons generated by the Anderson--Higgs mechanism \cite{GDMV1,GDMV2} in the surface region of the pin \cite{JY,JPY}.\footnote{This BEC shows no superfluidity or superconductivity because it is a charged ideal Bose gas.
This charged gas is stably confined due to the Anderson--Higgs mechanism; therefore, it requires no background charges.}
For its macroscopic radius $\approx25$ [$\mu$m] at temperature $\approx300$ [K], see Ref. \cite{GDMV2}.
In the Anderson--Higgs mechanism, the Goldstone boson is an {\it electric dipole phonon} in bound water molecules, which are electric dipoles in a resonant system with macroscopic {\it dynamical} order, in the surface region \cite{JY,JPY}.
Here, the {\it resonant system} refers to a two-lowest-rotational-level system of water molecules and the radiation field, which interact with each other resonantly.
The spatial translational degrees of freedom of a {\it bound water molecule} are suppressed and thus its position is approximately fixed in the resonant system \cite{JY,JPY}.
Owing to the non-vanishing effective electric charge $e^\ast$ of the condensed boson, this BEC can be manipulated electrically: for details, see footnote 2.

Specifically, the Hamiltonian of the resonant system in the approximation consists of three parts \cite{SWS}:
\begin{enumerate}
\item[(i)] The sum of the two-level intramolecular Hamiltonians of the bound water molecules.
\item[(ii)] The radiation field energy modeled by a single photon oscillator mode.
\item[(iii)] The linear coupling between the dipole moment per unit volume and the electric field modeled by the resonant photon creation and annihilation terms.
\end{enumerate}
Here, we neglect the sum of the screened (short-ranged) intermolecular Coulomb potentials (individually, the Coulomb interaction minus the conventional dipole-dipole interaction) between the neighboring bound water molecules \cite{SWS}.
Importantly, the self-energy of the electric dipolar system is thus reduced to the {\it block-diagonal} form, where each block represents an individual bound water molecule.
The self-energy in each block is incorporated into the two-level intramolecular Hamiltonian of the bound water molecule \cite{SWS}: the resonant system under the approximation is thus not equivalent to the {\it Dicke model} \cite{Dicke1,Dicke2}.

In the Heisenberg picture, a time-independent solution of the Heisenberg equations of motion defined for this Hamiltonian in terms of the canonical variables for the photon oscillator mode exhibits spontaneous breaking of the $U(1)$ rotational symmetry around the third axis for the {\it energy spin} \cite{AE} of the two-level water molecule \cite{JY,JPY}: this is the {\it dynamically ordered state} of the resonant system \cite{JY,JPY}.

Here, we show the consequent Anderson--Higgs mechanism in this system \cite{JY,GDMV1,GDMV2,JPY}.
The $U(1)$ transformation of the two-component energy spinor field of the two-level water molecule, $\Psi_s({\bx},t)$ (the subscript $s$ distinguishes the excited state and the ground state of the two-level water molecule), in the degenerate dynamically ordered state is generated by half of the third Pauli matrix $\widehat{\sigma}_3/2$.
Namely, we have the expression
\begin{equation}
\Psi_s(\bx,t)=\sqrt{N}_0\left(\begin{array}{c}e^{\frac{i}{2}\theta(\bx,t)}\alpha \\
e^{-\frac{i}{2}\theta(\bx,t)}\beta
\end{array}
\right)
\end{equation}
for complex numbers $\alpha$ and $\beta$ satisfying $|\alpha|^2+|\beta|^2=1$, the water-molecule number density $N_0$, and the Nambu--Goldstone field $\theta(\bx,t)$.
This energy spinor field with the non-zero effective charge $e^\ast$ has the non-relativistic minimal coupling to the radiation field.
Then, by redefining the gauge potential of the radiation field
\begin{equation}
A_\mu(\bx,t)\to A_\mu(\bx,t)-\frac{\hbar}{2e^\ast}\partial_\mu \theta(\bx,t)\;,
\end{equation}
the Nambu--Goldstone field $\theta(\bx,t)$ vanishes from the Hamiltonian, and instead of this, the radiation field acquires the effective mass.
The massive quanta of this redefined radiation field are the evanescent photons, and their BEC is system $A$.

In Fig. 1, we schematically show the whole system of a pin under consideration.

\begin{figure}[htbp]
\begin{center}
\includegraphics[width=0.525\hsize,bb=0 0 349 207]{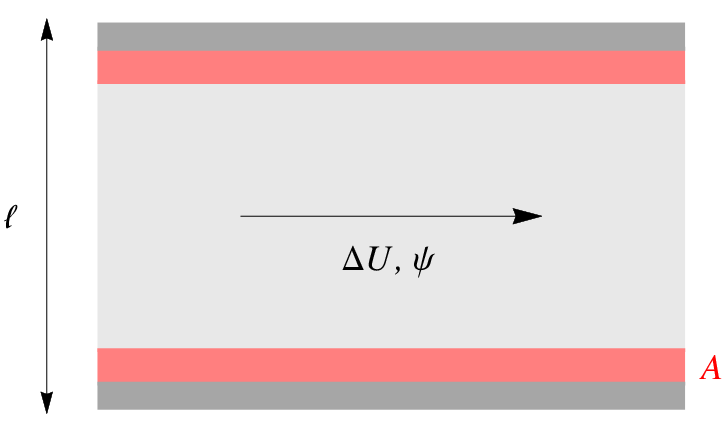}
\includegraphics[width=0.375\hsize,bb=0 0 260 234]{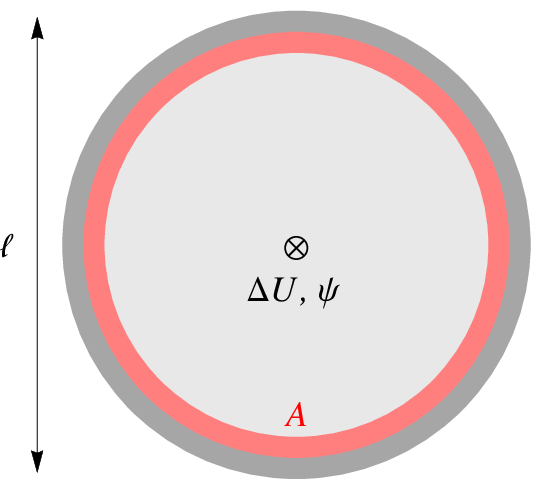}
\end{center}
\caption{Horizontal (left) and vertical (right) sections of a pin.
The gray, red, and light-gray regions (from outer to inner) show the hydrophilic membrane, the surface region, and the bulk water, respectively.
Applied voltage $\Delta U$ (system $\psi$) slopes to zero with a finite run length and propagates in the direction of the arrow (the $z$-direction).
System $A$ exists in the surface region.
The diameter $\ell$ of this tubule is shorter than the resonant wavelength $\approx400$ [$\mu$m] of the radiation from the two-level water molecule.}
\end{figure}

It is to be noted that, in the minimal-coupling model of the quantum electrodynamics of a resonant system, when we model the radiation field by a single mode (a uniform mode), the no-go theorem holds on the existence of a non-perturbative vacuum state (i.e., an equilibrium superradiant phase transition) \cite{APGMP1,APGMP2}.
Namely, in this single-mode model, the {\it ground state} of the resonant system is the perturbative vacuum state.
Here, the minimal-coupling model is transformed to the model in Ref. \cite{SWS} by a canonical transformation.

\subsection{Derivation of the von Neumann-type interaction}

In this subsection, we examine the state-reduction mechanism in a pin (system $\psi$).
Because the BEC (system $A$) consists of bosons with non-vanishing effective mass $m^\ast$, the center of mass of this macroscopic BEC has classical (i.e., mutually commuting) canonical variables $({\vec{\widehat{Q}}}_A,{\vec{{\widehat{P}}}}_A)$.
These classical canonical variables are the orbital superselection rule operators of system $A$ \cite{dEspagnat,Neumann,IJQI}.

Our objective is to identify the state reduction as a physical event with the non-zero electric potential $\Delta U=U_0$, which propagates down the tubule.
Here, the electric potential $\Delta U$ sloping to zero with a finite run length is assumed to be binary, that is, $\Delta U=0$ or $U_0$.
We denote by $\widehat{{\mathfrak M}}_\psi$ the binary meter variable of system $\psi$, which represents this electric potential.
Specifically,
\begin{eqnarray}
{\mathfrak M}_\psi=\left\{\begin{array}{ccc}0&& (\Delta U=0) \\  && \\ 1&&(\Delta U=U_0)\end{array}\right.\;.\label{eq:O}
\end{eqnarray}
When the electric potential $U_0$ is applied, the non-zero external Coulomb force $e^\ast U_0/l_r$ acts on the condensed boson (see footnote 2).
Here, $l_r$ denotes the run length of the electric potential sloping to zero in the $z$-direction (the direction of the electric-potential propagation down the tubule).
We assume this slope to be uniform.
In the zeroth approximation, when the speed of the BEC along the $z$-axis is slower than the speed of sound in liquid water, the BEC is acted on by the other classical force---that is, the Coulomb collisions of the BEC, accelerated along the $z$-axis, with the bulk water molecules---which has a constant of proportionality $-k_dm^\ast$ in a linear ($d=1$) or quadratic ($d=2$) relation with respect to the velocity $V^z_A$ of the BEC in the $z$-direction (see footnote 2).

We analyze this setting classical mechanically.
The simplified model has the following equation of motion:
\begin{eqnarray}
m^\ast \frac{dV^z_A}{dt}=m^\ast a_C-k_d m^\ast \left(V^z_A\right)^d\;,
\end{eqnarray}
where $d=1$ or $2$ and
\begin{eqnarray}
a_C=\frac{1}{m^\ast}\left(\frac{e^\ast U_0}{l_r}\right)\;.
\end{eqnarray}
After the relaxation time (collision time) $\tau_d=1/k_1$ ($d=1$), $1/\sqrt{a_C k_2}$ ($d=2$) elapses on a quantum mechanical time scale, $V^z_A$ would relax to the terminal velocity
\begin{eqnarray}
V^z_{A,\infty,d}&=&\left(\frac{a_C}{k_d}\right)^{1/d} \\
&=&a_C \tau_d\;,
\end{eqnarray}
which is independent from the initial velocity.

Alternatively, we denote the terminal velocity by
\begin{equation}
\Lambda=V^z_{A,\infty,d}\;.
\end{equation}
By this constant velocity, the BEC is spatially displaced along the $z$-axis while the electric potential $U_0$ is applied.
This is because the electric potential $U_0$ propagates down the tubule on a classical mechanical time scale.
Finally, we incorporate the orbital superselection rule into system $A$.
Then, we obtain the von Neumann-type interaction between systems $\psi$ and $A$, that is,
\begin{equation}
\widehat{H}_{\psi{\text{-}} A}=\Lambda \widehat{{\mathfrak M}}_\psi \otimes \widehat{P}^z_A\label{eq:H}
\end{equation}
as the quantum mechanical interaction Hamiltonian between systems $\psi$ and $A$ approximately.

According to Ref. \cite{EPL3}, this interaction Hamiltonian between systems $\psi$ and $A$ realizes the state-reduction mechanism in system $\psi$ in the interaction picture.

\section{Conclusion}

In a computer with the von Neumann architecture, we incorporate the binary projection hypothesis (a state-reduction mechanism) into each of the pins (system $\psi$) on the CPU; here, each of the pins realizes the von Neumann-type interaction with a BEC (system $A$) generated by the Anderson--Higgs mechanism.
We use projective quantum measurement as a classical measurement (reduction of ignorance).

In the setting for the state-reduction mechanism, there are two crucial points.
The first is that system $A$ is a macroscopic and field-theoretical BEC, which is both massive and electrically charged \cite{JY}.
The second is that the binary quantum eigenstate of the meter variable of system $\psi$ represents the binary value of the electric potential $\Delta U=0$, $U_0$ during the process.
In this mechanism, after completion of the required quantum decoherence process \cite{JSTAT2}, the electric potential serves as a binary switch for the state reduction of system $\psi$.
Note that, in this state-reduction mechanism, our entire framework falls within quantum electrodynamics in the presence of the orbital superselection rule.

There are two novel elements in our architecture compared with the von Neumann architecture.
The first is the incorporation of the coarse-graining into the registers in the CPU in the computation cycles for binary data from the registers to the ALU, and back to the registers.
With this coarse-graining incorporated, the computer can acquire positive information by the state reduction.
The second is, as summarized above, the incorporation of the state-reduction mechanism into the pins on the CPU, which are located between the main memory and the registers in the CPU, in the write (respectively, read) operation of a binary datum into (respectively, from) the main memory.

Our state-reduction mechanism in the CPU establishes the information-acquiring von Neumann architecture of a computer.
As a consequence of this architecture, information-acquiring artificial intelligence, whose logic is based on inductive inference implemented on the computer as software\footnote{In universal inductive inference \cite{LV}, the posterior probability distribution over generating programs (laws) conditioned on the observed data is computed by the ALU and then stored in the main memory, in practical spatio-temporally resource-bounded models.
This distribution is sequentially updated via Bayes' theorem.} \cite{Aggarwal,Hutter}, can also be established as an equivalent to the observer in quantum mechanics.
Besides its functionality property, it is able to gain knowledge of events and store it in the main memory as information.
For the observer, this ability is the fundamental postulate in information theory \cite{Shannon}.
Finally, we note again that, in classical measurement, the entropy of events before an event reading has the classical statistical mechanical origin.


\begin{thebibliography}{99}
\bibitem{dEspagnat}B. d'Espagnat,
{\it Conceptual Foundations of Quantum Mechanics}, 2nd edn. (W. A. Benjamin, Reading, Massachusetts, 1976).
\bibitem{Neumann}J. von Neumann,
{\it Mathematical Foundations of Quantum Mechanics} (Princeton University Press, Princeton, NJ, 1955).
\bibitem{Luders}G. L$\ddot{{\rm u}}$ders,
{\it Ann. Phys. (Leipzig)} {\bf 8} (1951) 322.
\bibitem{EPL3}E. Konishi,
{\it Europhys. Lett.} {\bf 136} (2021) 10004.
\bibitem{IJQI}E. Konishi,
{\it Int. J. Quantum Inf.} {\bf 22} (2024) 2450033.
\bibitem{NS}N. Nisan and S. Schocken,
{\it The Elements of Computing Systems: Building a Modern Computer from First Principles}, 2nd edn. (The MIT Press, Cambridge, Massachusetts, 2021).
\bibitem{PH}D. A. Patterson and J. L. Hennessy,
{\it Computer Organization and Design MIPS Edition}, 6th edn. (Elsevier, Amsterdam, 2021).
\bibitem{GPV}E. Del Giudice, G. Preparata and G. Vitiello,
{\it Phys. Rev. Lett.} {\bf 61} (1988) 1085.
\bibitem{Talkner}P. Talkner, E. Lutz and P. H$\ddot{{\rm a}}$nggi,
{\it Phys. Rev. E} {\bf 75} (2007) 050102(R).
\bibitem{MT}Y. Morikuni and H. Tasaki,
{\it J. Stat. Phys.} {\bf 143} (2011) 1.
\bibitem{JSTAT2}E. Konishi,
{\it J. Stat. Mech.: Theor. Exp.} {\bf 2019}(1) (2019) 019501.
\bibitem{Matsumoto1}H. Matsumoto, G. Oberlechner, M. Umezawa and H. Umezawa, {\it J. Math. Phys.} {\bf 20} (1979) 2088.
\bibitem{Matsumoto2}H. Matsumoto, G. Semenoff, H. Umezawa and M. Umezawa, {\it J. Math. Phys.} {\bf 21} (1980) 1761.
\bibitem{Matsumoto3}G. Semenoff, H. Matsumoto and H. Umezawa, {\it J. Math. Phys.} {\bf 22} (1981) 2208.
\bibitem{Matsumoto4}H. Matsumoto, N. J. Papastamatiou, G. Semenoff and H. Umezawa, {\it Phys. Rev. D} {\bf 24} (1981) 406.
\bibitem{Umezawa}H. Umezawa, {\it Advanced Field Theory: Micro, Macro and Thermal Physics} (American Institute of Physics, New York, 1993).
\bibitem{NC}M. A. Nielsen and I. L. Chuang, {\it Quantum Computation and Quantum Information} (Cambridge University Press, Cambridge, 2000).
\bibitem{JY}M. Jibu and K. Yasue,
{\it Informatica} {\bf 21} (1997) 471.
\bibitem{GDMV1}E. Del Giudice, S. Doglia, M. Milani and G. Vitiello,
{\it Nucl. Phys. B} {\bf 275} (1986) 185.
\bibitem{GDMV2}E. Del Giudice, S. Doglia, M. Milani and G. Vitiello,
{\it Phys. Scr.} {\bf 38} (1988) 505.
\bibitem{JPY}M. Jibu, K. H. Pribram and K. Yasue,
{\it Int. J. Mod. Phys. B} {\bf 10} (1996) 1735.
\bibitem{SWS}S. Sivasubramanian, A. Widom and Y. N. Srivastava, {\it Physica A} {\bf 301} (2001) 241.
\bibitem{Dicke1}R. H. Dicke, {\it Phys. Rev.} {\bf 93} (1954) 99.
\bibitem{Dicke2}M. M. Roses and E. G. Dalla Torre, {\it PLoS ONE} {\bf 15} (2020) e0235197.
\bibitem{AE}L. Allen and J. H. Eberly, {\it Optical Resonance and Two-Level Atoms} (Wiley, New York, 1975).
\bibitem{APGMP1}G. M. Andolina, F. M. D. Pellegrino, V. Giovannetti, A. H. MacDonald and M. Polini, {\it Phys. Rev. B} {\bf 100} (2019) 121109(R).
\bibitem{APGMP2}G. M. Andolina, F. M. D. Pellegrino, V. Giovannetti, A. H. MacDonald and M. Polini, {\it Phys. Rev. B} {\bf 102} (2020) 125137.
\bibitem{LV}M. Li and P. M. B. Vit$\acute{{\rm a}}$nyi,
{\it An Introduction to Kolmogorov Complexity and Its Applications}, 4th edn. (Springer Nature, Cham, Switzerland, 2019).
\bibitem{Aggarwal}C. C. Aggarwal,
{\it Artificial Intelligence: A Textbook} (Springer Nature, Cham, Switzerland, 2021).
\bibitem{Hutter}M. Hutter,
{\it Universal Artificial Intelligence: Sequential Decisions Based on Algorithmic Probability} (Springer-Verlag, Berlin, 2005).
\bibitem{Shannon}C. E. Shannon,
{\it Bell Syst. Tech. J.} {\bf 27} (1948) 379.
\end{thebibliography}
\end{document}